\documentclass[aps,prl,twocolumn,groupedaddress,showpacs,floatfix]{revtex4}
\usepackage{graphics}
\usepackage{epsfig}
\usepackage{amsmath}
\newcommand{\one}{$^{17}$O(p,$\alpha$)$^{14}$N}
\newcommand{\two}{$^{18}$O(p,$\alpha$)$^{15}$N}

\newcommand{\og}{$\omega\gamma$}

\usepackage{dcolumn}% Align table columns on decimal point
\begin{document}
\title{Improved direct measurement of the 64.5 keV resonance strength in the $^{17}$O(p,$\alpha$)$^{14}$N reaction at LUNA}

\author{
	C.G.\,Bruno,$^{1,*}$
	D.A.\,Scott,$^1$ 
	M.\,Aliotta,$^{1,*}$	
	A.\,Formicola,$^2$
	A.\,Best,$^3$	
	A.\,Boeltzig,$^4$
	D.\,Bemmerer,$^5$	
	C.\,Broggini,$^6$		
	A.\,Caciolli,$^7$ 
	F.\,Cavanna,$^8$
	G.F.\,Ciani,$^4$
	P.\,Corvisiero,$^8$	
	T.\, Davinson,$^1$
	R.\,Depalo,$^7$
	A.\,Di Leva,$^3$
	Z.\,Elekes,$^9$
	F.\,Ferraro,$^8$	
	Zs.\,F\"ul\"op,$^9$	
	G.\,Gervino,$^{10}$
	A.\,Guglielmetti,$^{11}$
	C.\,Gustavino,$^{12}$
	Gy.\,Gy\"urky,$^9$
	G.\,Imbriani,$^3$
	M.\,Junker,$^2$
	R.\,Menegazzo,$^6$
	V.\,Mossa,$^{13}$
	F.R.\,Pantaleo,$^{13}$
	D.\,Piatti,$^7$
	P.\,Prati,$^8$
	E.\,Somorjai,$^9$
	O.\,Straniero,$^{14}$
	F.\,Strieder,$^{15}$
	T.\,Sz\"ucs,$^5$
	M.P.\,Tak\'acs,$^5$
	D.\,Trezzi$^{11}$\\
	(The LUNA Collaboration)
	}

\affiliation{$^1$SUPA, School of Physics and Astronomy, University of Edinburgh, EH9 3FD Edinburgh, United Kingdom} % 1
\affiliation{	$^2$INFN, Laboratori Nazionali del Gran Sasso (LNGS), 67100 Assergi, Italy} % 2
\affiliation{	$^3$Universit\`a di Napoli ``Federico II'', and INFN, Sezione di Napoli, 80126 Napoli, Italy} % 3
\affiliation{	$^4$Gran Sasso Science Institute, INFN, Viale F. Crispi 7, 67100 L'Aquila, Italy} % 3
\affiliation{	$^5$Helmholtz-Zentrum Dresden-Rossendorf, Bautzner Landstr. 400, 01328 Dresden, Germany} % 4
\affiliation{	$^6$INFN, Sezione di Padova, Via F. Marzolo 8, 35131 Padova, Italy} % 4
\affiliation{	$^7$Universit\`a degli Studi di Padova and INFN, Sezione di Padova, Via F. Marzolo 8, 35131 Padova, Italy} % 5
\affiliation{	$^8$Universit\`a degli Studi di Genova and INFN, Sezione di Genova, Via Dodecaneso 33, 16146 Genova, Italy} % 5
\affiliation{	$^9$Institute for Nuclear Research (MTA ATOMKI), PO Box 51, HU-4001 Debrecen, Hungary} % 8
\affiliation{	$^{10}$Universit\`a degli Studi di Torino and INFN, Sezione di Torino, Via P. Giuria 1, 10125 Torino, Italy} % 9
\affiliation{	$^{11}$Universit\`a degli Studi di Milano and INFN, Sezione di Milano, Via G. Celoria 16, 20133 Milano, Italy} % 5
\affiliation{$^{12}$INFN, Sezione di Roma La Sapienza, Piazzale A. Moro 2, 00185 Roma, Italy}
\affiliation{	$^{13}$Universit\`a degli Studi di Bari e INFN, Sezione di Bari, 70125 Bari, Italy} % 5
\affiliation{	$^{14}$Osservatorio Astronomico di Collurania, Teramo, and INFN, Sezione di Napoli, 80126 Napoli, Italy} % 11
\affiliation{	$^{15}$South Dakota School of Mines, 501 E. Saint Joseph St., SD 57701 USA} % 10

\date{\today}

\begin{abstract}
The \one\ reaction plays a key role in various astrophysical scenarios, from asymptotic giant branch stars to classical novae. It affects the synthesis of rare isotopes such as $^{17}$O and $^{18}$F, which can provide constraints on astrophysical models. A new direct determination of the $E_{\rm R}~=~64.5$~keV resonance strength performed at the Laboratory for Underground Nuclear Astrophysics accelerator has led to the most accurate value to date, $\omega\gamma = 10.0 \pm 1.4_{\rm stat} \pm 0.7_{\rm syst}$~neV, thanks to a significant background reduction underground and generally improved experimental conditions. 
The (bare) proton partial width of the corresponding state at $E_{\rm x} = 5672$~keV in $^{18}$F is $\Gamma_{\rm p} = 35 \pm 5_{\rm stat} \pm 3_{\rm syst}$~neV. This width is about a factor of 2 higher than previously estimated thus leading to a factor of 2 increase in the \one\ reaction rate at astrophysical temperatures relevant to shell hydrogen-burning in red giant and asymptotic giant branch stars. The new rate implies lower $^{17}$O/$^{16}$O ratios, with important implications on the interpretation of astrophysical observables from these stars. 
\end{abstract}


\maketitle

\label{sec:introduction}

% introduction
Measurements of the C, N and O isotopic ratios in the atmosphere of red giant branch (RGB), asymptotic giant branch (AGB) stars, and in meteoritic stardust grains originating from AGB stars \cite{nittler1997}, can be used to test the efficiency of deep mixing processes, such as those due to convection, rotational instabilities, magnetic buoyancy, thermohaline circulation, and gravity waves \cite{el-eid1994,boothroyd1999,lebzelter2015}. Indeed, variations in these isotopic ratios are expected to occur at the surface of a giant star when the mixing extends down to the stellar interior where the H burning takes place. Thus, a knowledge of the rates of reactions belonging to the CNO cycle is required. In particular, an accurate and precise knowledge of the \one\ reaction can place important constraints on the $^{17}$O/$^{16}$O abundance ratio predicted by different astrophysical models \cite{nittler1997}. An improved measurement of the \one\ reaction rate may also help shed some light on the peculiar composition of some pre-solar grains \cite{lugaro2007}.

Over a wide range of astrophysical temperatures relevant to quiescent- ($T = 0.03-0.1$~GK) and explosive- ({\em e.g.}, up to $T = 0.4$~GK in novae) H burning, the reaction rate of \one\ is dominated by two isolated and narrow resonances, respectively at $E_{\rm R}$ = 183 and 64.5~keV (in the center-of-mass system). However, while measurements of the 183~keV resonance strength \cite{chafa2005,chafa2006,chafa2007,moazen2007,newton2007,newton2010} have led to values that are in good agreement with one another and yield a weighted average of $\omega \gamma = (1.66\pm0.10)\times10^{-6}$~eV \cite{sergi2010}, the strength of the 64.5~keV resonance is still uncertain. An early attempt at its direct measurement \cite{berheide1992} reported an upper limit ($\omega\gamma<0.8$~neV), later superseded ($\omega\gamma \le 22.0$~neV) by an unpublished re-analysis \cite{niemeyer1996} of the same data. A subsequent study \cite{blackmon1995}, under infinitely thick-target conditions, reported a value of $\Gamma_{\rm p} = 22\pm3$~neV for the proton partial width, from which a strength $\omega\gamma = 5.5^{+1.8}_{-1.5}$~neV (value adopted in NACRE compilation \cite{angulo1999}) was inferred. Later re-analyses \cite{hannam1999,iliadis2016}, however, led to slightly different values, including $\Gamma_{\rm p} = 19\pm2$~neV \cite{iliadis2016}, as adopted in the STARLIB compilation \cite{starlib}.
Indirect investigations by Sergi {\em et al.} \cite{sergi2010,sergi2015} based on the Trojan Horse Method (THM) arrived at a bare ({\em i.e.}, free from electron screening effects) weighted average \og\ $ = 3.42\pm{0.60}$~neV. 
The most recent determination of the \one\ reaction rate is reported in Buckner {\em et al.} \cite{buckner2015} where an increase of 30\% compared to the rate of Iliadis {\em et al.} \cite{iliadis2010} is ascribed to the influence of a -2~keV sub-threshold state in $^{18}$F.   

In this Letter, we present the results of an accurate measurement of the $E_{\rm R} = 64.5$~keV resonance strength in the \one\ reaction and provide a new recommended astrophysical reaction rate. This measurement forms part of a scientific program of H-burning reaction studies \cite{scott2012,cavanna2015} at the Laboratory for Underground Nuclear Astrophysics (LUNA)\cite{broggini2010,costantini2009}.

The experiment was performed at the LUNA 400~kV accelerator \cite{formicola2003} of Laboratori Nazionali del Gran Sasso (LNGS) of INFN, Italy, using intense ($\approx~150~\mu$A), low-energy proton beams on Ta$_2$O$_5$ targets \cite{caciolli2012} isotopically enriched (80-85\% ca.) in $^{17}$O  ($\Delta E_{\rm lab} \approx 6$~keV at the resonant energy, corresponding to a proton beam energy in the laboratory frame $E_{\rm p} = 70$~keV). 
Targets were also doped with a small amount (a few percent) of $^{18}$O for calibration and setup commissioning purposes \cite{bruno2015}.
We typically replaced targets after about 20C of accumulated charge to keep target degradation to about 10\% at most, corresponding to a reduction in target thickness of less than 1~keV \cite{bruno2015} and thus ensuring thick-target conditions throughout the measurement (see below).

Because of the low bombarding energies, the kinematics of the emitted alpha particles is essentially determined by the Q-value (Q$ = 1.192$~MeV) of the \one\ reaction. Thus, an experimental setup for high-efficiency, low-energy alpha-particle detection was developed and commissioned \cite{bruno2015}. Briefly, the setup consisted of an array of silicon detectors ($300-700~\mu$m thick), with a total efficiency at backward angles (135.0$^{\circ}$ and 102.5$^{\circ}$ to the beam axis) and at a distance of 6~cm from the target of about 10\%, as obtained by Monte Carlo simulations \cite{bruno2015}. Aluminized Mylar foils (2.4~$\mu$m thick) were mounted in front of each detector to protect them from elastically scattered protons. Calculated energy loss (TRIM, \cite{srim}) for alpha particles from the reaction was about $750$~keV (in the lab). Signals from the detectors were processed with standard electronics and acquired in list-mode (MIDAS \cite{midas}), with a trigger given by the logical OR of the silicon detectors' signals. The measurements took place for an overall accumulated charge on target of 75~C off-resonance ($E_{\rm p} = 65$~keV) and 137~C on-resonance ($E_{\rm p} = 71.5$~keV). Background runs (no beam on target) for a total of 217~hours were also acquired. 

% method

The $E_{\rm R} = 64.5$~keV resonance strength $\omega \gamma$ can be determined directly from experimental yields using a thick-target approach \cite{iliadis-book}:
\begin{equation}
\omega \gamma = \frac{2}{\lambda^2}\epsilon_{\rm eff}\frac{N_\alpha}{\eta W N_{\rm p}}=\frac{2}{\lambda^2}\epsilon_{\rm eff}\frac{N_\alpha e}{\eta W Q}
\label{eq:omega}
\end{equation}
where $N_\alpha$ and $N_{\rm p} ( = Q/e)$ represent the number of detected alpha particles and incident protons, respectively (with $e$ and $Q$ elementary and total accumulated charge, respectively); $\epsilon_{\rm eff}$ is the effective stopping power in the center-of-mass frame; $W=W(\theta)$ is the angular distribution at detector laboratory angle $\theta$; $\eta$ the detection efficiency; and $\lambda$ the de Broglie wavelength at the center-of-mass resonant energy.

%definition of ROI
\begin{figure}[t!]
\includegraphics[width=\columnwidth]{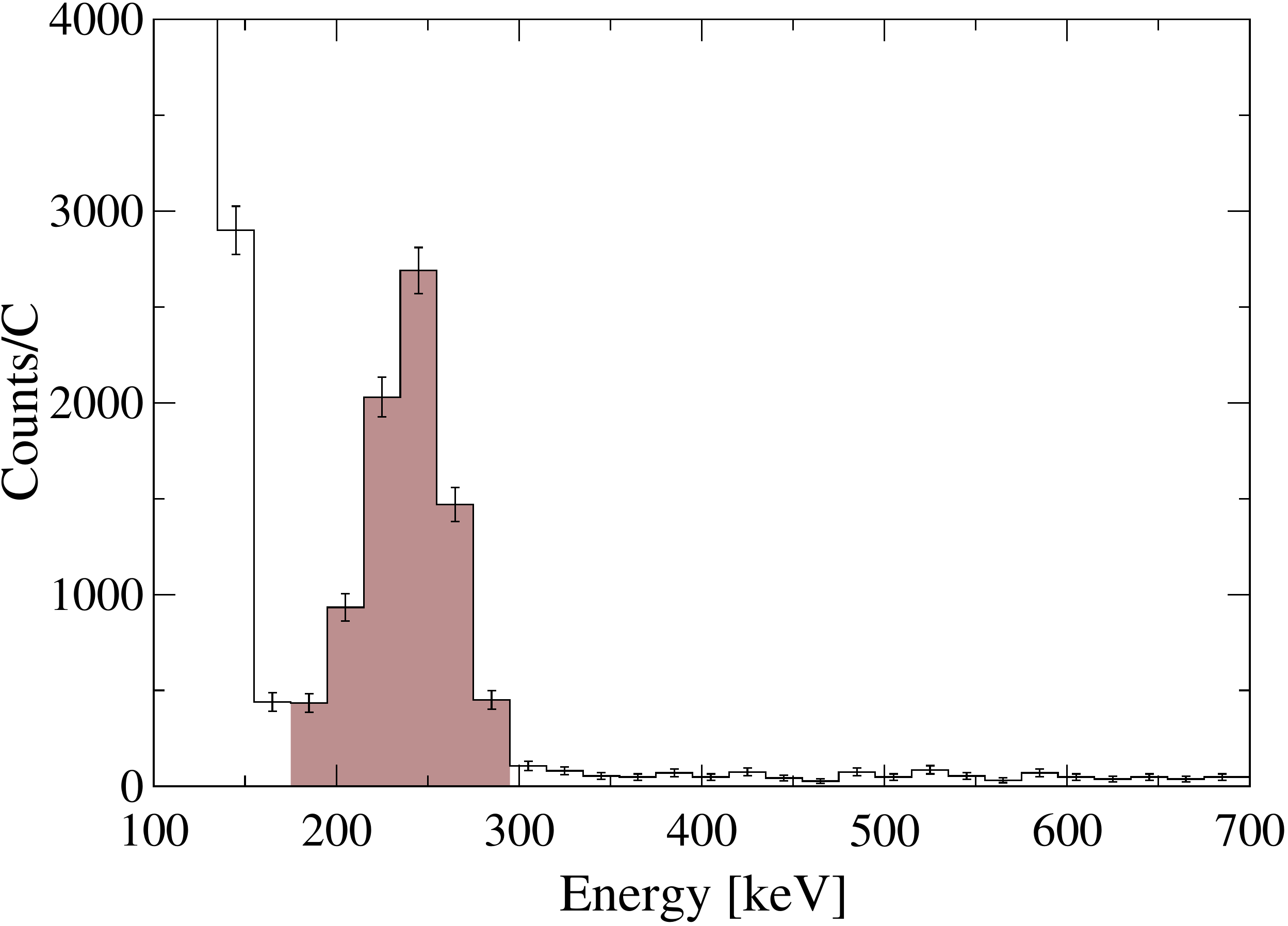}
\caption{Typical $\alpha$-particle spectrum from the $E_{\rm R} =183$~keV resonance at $E_{\rm p}=193$~keV. The position and width of the alpha particle peak (shaded area) is used to help define the ROI for the $E_{\rm R} = 64.5$~keV resonance alpha particles.  Error bars correspond to statistical uncertainties.}
\label{fig:193}
\end{figure}

The expected alpha particle yield from the $E_{\rm R}~=~64.5$~keV resonance is extremely low (about a thousand times lower than for the $E_{\rm R}~=~183$~keV resonance). It was thus critical to define the relevant region of interest (ROI) as accurately as possible.
To this end, we exploited the stronger alpha peak from the $E_{\rm R}=183$~keV ($E_{\rm p}=193$~keV) resonance, not known at the time of previous direct studies. Based on the measured location and total width of this peak  (Fig. \ref{fig:193}) in each detector ($E_\alpha \approx 250$~keV and $\Delta E_\alpha \approx 100$~keV, respectively, after passing through the aluminized Mylar foils \cite{srim}), we defined the ROI for $E_{\rm p} = 70$~keV resonance as being approximately 100~keV wide and centered around about 200~keV (with actual values varying detector by detector) to take into account a small energy shift ($\approx$ 50 keV) due to the combined effect of the lower beam energy (70 keV {\em cf.} 193 keV) and the slightly lower $\alpha$-particle energy loss through the foils.  Energy loss calculations were done using TRIM \cite{srim} and the values reported here are in the laboratory system.
We also checked that no shifts in the energy calibration \footnote{\label{note:calib}An effective detector calibration was obtained through a minimisation procedure as described in \cite{bruno2015}. Briefly, we optimised calibration parameters and foils thickness, detector by detector, to simultaneously reproduce the energy of the alpha peaks from the $E_{\rm p}=193$~keV resonance in \one\ and the $E_{\rm p}=151$~keV resonance in \two. See refs. \cite{bruno2015, bruno2016} for further details.} occurred during the whole data taking campaign, which might have affected the ROIs. This was achieved by daily monitoring the position of the alpha-peak centroid from the $E_{\rm p} = 151$~keV resonance in the \two\ reaction \cite{bruno2015}. 

% checks on ROI - quality data

To further ensure the quality and reliability of our data, additional selection criteria were employed: data were divided in fragments of 20 minutes each, and rates in the ROI and in adjacent energy regions were monitored for stability over time using a Maximum Likelihood approach. We selectively discarded data with rates five or more standard deviations away from average values as well as data with known experimental issues. Approximately 30\% of all data were rejected following this procedure. Target degradation due to intense beam bombarding (resulting in reduced target thickness and thick-target yield by about 0.5\% per Coulomb in each case \cite{bruno2015,bruno2016}) was properly taken into account in the data analysis. Details of target stoichiometry measurements are given in Ref. \cite{caciolli2012}.

% data analysis & results

\begin{figure}
	\includegraphics[width=\columnwidth]{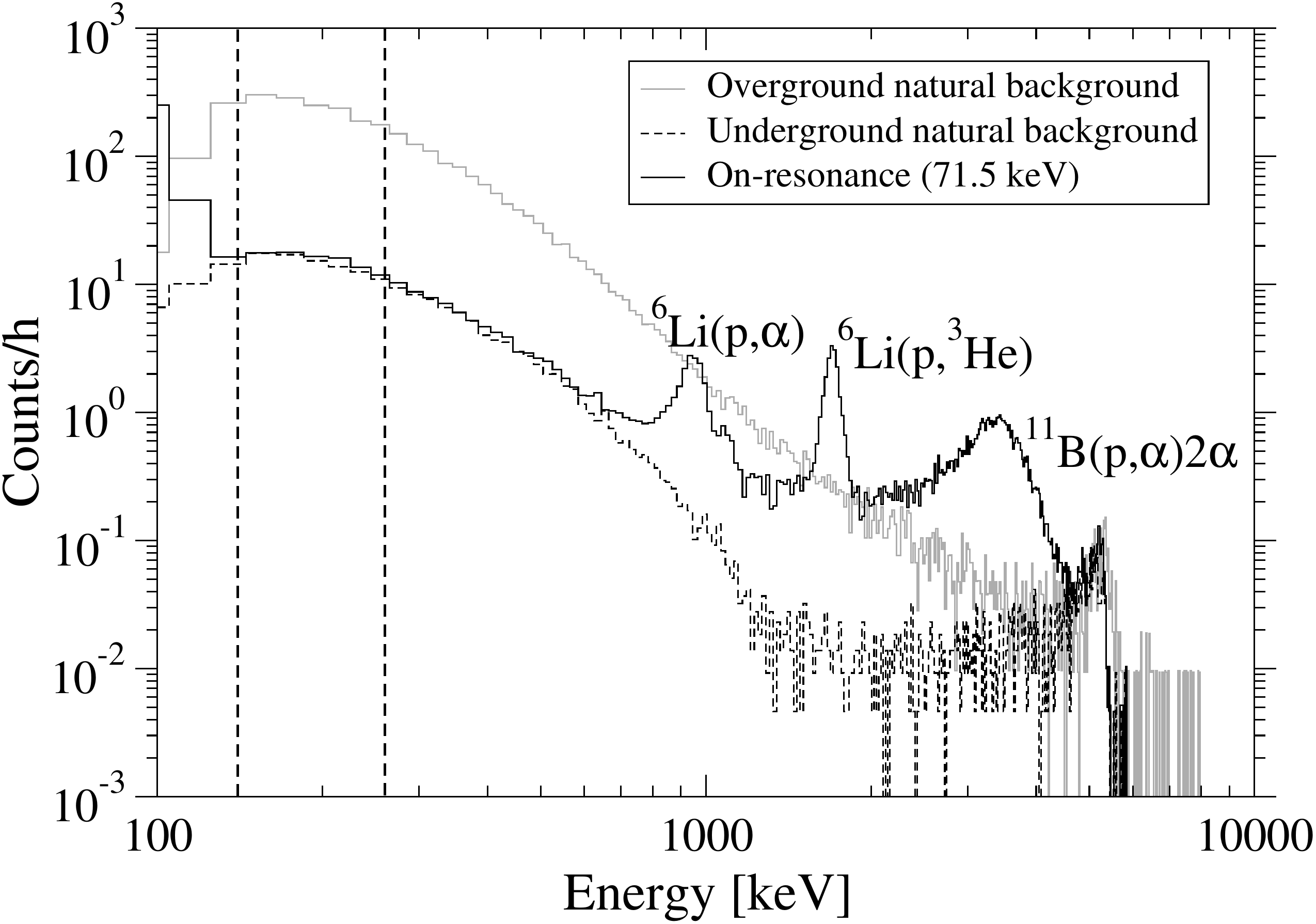}
	\caption{Overlay of background spectra taken overground- (grey line) and underground (dashed line) showing a factor of 15 background reduction in the region of interest (vertical lines) of expected alpha particles from the $E_{\rm R} = 64.5$~keV resonance in \one. Also shown is the full energy spectrum (black line) taken on resonance at $E_{\rm p}=71.5$~keV. Peaks arise from beam-induced reactions on target contaminants as labeled. The peak around 5~MeV is ascribed to intrinsic alpha activity in the silicon detectors \cite{bruno2015}.} 
	\label{fig:fullenergy}
\end{figure}

Figure \ref{fig:fullenergy} shows an overlay of background spectra taken over ground and underground, as well as the full energy spectrum taken on resonance ($E_{\rm p}=71.5$~keV) and summed over all detectors (all spectra have been properly normalised in time). Note a factor of 15 reduction in underground background in the ROI (vertical dashed lines), demonstrating a crucial advantage of underground measurements for charged particle detection at these energies. 
The exponential-like feature at low $E \leq 100$~keV in the on-resonance spectrum is due to a combination of natural background and electronic noise. Peaks at higher energies are due to $^3$He- and alpha-particles from beam-induced background reactions on $^6$Li and $^{11}$B contaminants in the target. No evidence was found for reactions on $^7$Li and $^{10}$B. To assess any potential contribution of beam-induced background peaks in the region of interest, we employed a correlation analysis between counts in the ROI and counts in the background peaks, detector by detector. We found no evidence of any contributions from such reactions to events in the ROI. 
 
In order to extract the net counts of alpha particles, $N_\alpha$, in the ROI of the $E_{\rm R} = 64.5$~keV resonance, the following subtraction procedure should be applied:
\begin{itemize}
\item[1)] normalise all spectra (on-resonance, off-resonance, and natural background) to time;
\item[2)] subtract the natural background from the on- and off-resonance spectra (bin by bin);
\item[3)] normalise the resulting on- and off-resonance spectra in charge;
\item[4)] subtract the charge-normalised off-resonance spectrum from the on-resonance one (bin by bin);
\item[5)] integrate the counts in the ROI.
\end{itemize}
\begin{figure}[htp]
  \includegraphics[width=\columnwidth]{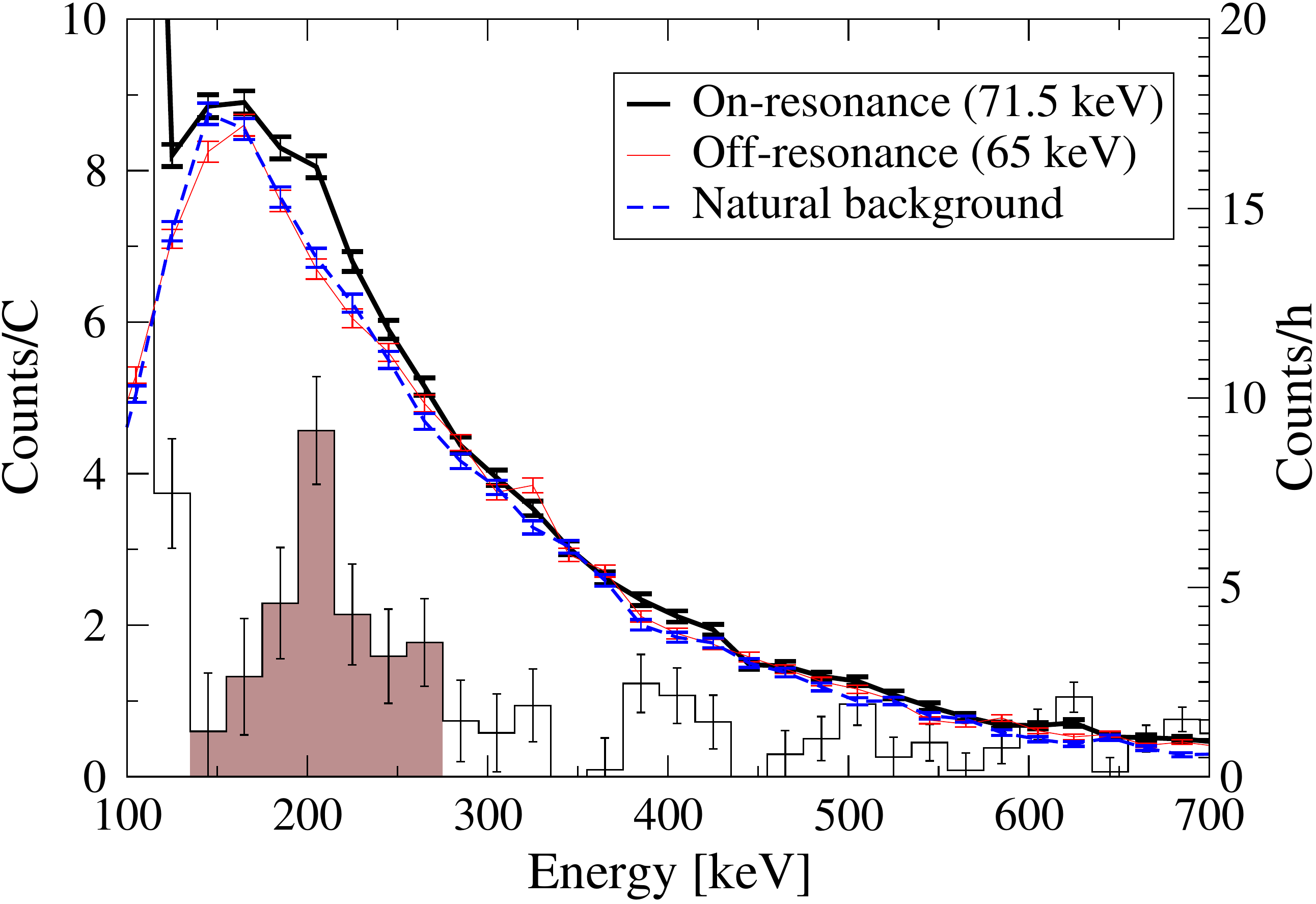}
\caption{(Color online) Overlay of time-normalised on-resonance ($E_{\rm p}=71.5$~keV), off-resonance ($E_{\rm p}=65$~keV) and natural background spectra in counts/h (lines are to guide to eye). Also shown is the histogram (in counts/C) obtained after a bin-by-bin subtraction of the time-normalised natural background spectrum from the on-resonance one. The shaded peak corresponds to the region of interest of the alpha particles from the 64.5~keV resonance. Note the different y-axes.}
\label{fig:70}
\end{figure}

Time-normalised spectra (on-resonance, off-resonance, and natural background) are shown in Fig. \ref{fig:70}, with y-axis on the right hand-side. Also shown is the histogram obtained from a bin-by-bin subtraction of the natural background spectrum from the on-resonance spectrum in counts/C (y-axis on the left-hand side). The shaded region represents the ROI of the expected alpha particles from the $E_{\rm R} = 64.5$~keV resonance in \one. Note a factor of about 600 reduction in count rate compared to that from the $E_{\rm R} = 183$~keV resonance (Fig. \ref{fig:193}).

Numerical values of the spectra subtraction procedure are reported in Table \ref{tab:counts}. Here, $N_\alpha$ represents the integral of counts in the ROI (on-resonance spectrum) after subtracting, bin by bin, the time-normalised background spectrum (beam off run); NB gives natural background counts in the ROI (after time normalisation to the on-resonance spectrum); and BIB represents the beam-induced background counts in the ROI (off-resonance run) after subtraction of the time-normalised natural background spectrum. Note that the BIB value is negative, though consistent with zero, because the off-resonance spectrum and the natural background spectrum are essentially indistinguishable from one another (Fig. \ref{fig:70}). 
In order to avoid subtracting a negative number (beam-induced background) from $N_\alpha$ (point 4)), which would lead to an artificially higher resonance strength, 
we adopted a second more sophisticated analysis to determine the net alpha-particle counts in the ROI, $N_\alpha$, based on the Maximum Likelihood approach. 

Following the procedure described in sec 5.1 of Ref. \cite{cowan2011}, we used a likelihood function $L$ given by the product of three Poisson distributions:
\begin{multline}
L = P_{\rm on}(\mu+b+\nu;n)\,P_{\rm bkg}(\tau_bb;m)\,P_{\rm off}(\tau_ob+\xi \nu;l) 
\label{eq:ml}
\end{multline}
where $n$, $m$ and $l$ are respectively the total number of events observed in the ROI for on-resonance, off-resonance and background runs; and $\mu$, $b$, $\nu$ are respectively the estimators for the signal events, natural background, and beam-induced background events. These  estimators are not known {\em a priori}. The terms $\tau_b=t_{\rm bkg}/t_{\rm on}$ and $\tau_o=t_{\rm off}/t_{\rm on}$ represent (live) time normalisation factors for background and off-resonance runs, respectively; $\xi=Q_{\rm off}/Q_{\rm on}$ is the charge normalisation factor for off-resonance runs.
We maximized Eq. \ref{eq:ml} numerically with the condition $\nu \ge$ 0 to obtain the best values for $\mu$, $b$ and $\nu$, corresponding to the maximum likelihood of the observation. For ease of computation, uncertainties in these best values were estimated by a Monte Carlo simulation technique described in Refs. \cite{cowan2011,cowan-book}. Results are in excellent agreement with those obtained from the background subtraction approach (Table \ref{tab:counts}). 
\begin{table}[tb]
\caption{Net total counts (integral over the ROI) obtained using the spectra-subtraction method and the Maximum Likelihood approach (see text for details). }
	\begin{tabular}{ l c c }
	\hline
	
	Event type & Subtraction & Maximum  \\ 
	~ & Method & Likelihood \\ 
	\hline
	
$N_{\rm \alpha}$ (on-resonance spectrum) & 1222 $\pm$ 165& 1257 $\pm$ 178 \\
NB$^{a)}$ (beam-off spectrum) & 13814~$\pm$~111 & 13779 $\pm$ 98 \\
BIB$^{b)}$ (off-resonance spectrum) & -109 $\pm$ 194 & 0 $^{+55^{c)}}_{-0}$ \\
	\hline
 \footnotesize	$^{a)}$ NB = natural background\\
\footnotesize	$^{b)}$ BIB = beam-induced background\\
\footnotesize	$^{c)}$ At the 68\% confidence level
	\end{tabular}
\label{tab:counts}
\end{table}
We opt to use results from this Maximum Likelihood analysis to determine the $E_{\rm R} = 64.5$~keV resonance strength value, obtaining $\omega\gamma = 10.0 \pm 1.4_{\rm stat} \pm 0.7_{\rm syst}$~neV, with sources of uncertainty listed in Table \ref{tab:unc}. Note that an angular distribution $W(\theta) \approx 1$  was used as we inferred from an R-matrix calculation with AZURE2 \cite{azure} based on known properties of this resonance and validated on the $E_{\rm R} = 183$~keV resonance against data by \cite{chafa2007}. The effective stopping power $\epsilon_{\rm eff}$ (Eq. \ref{eq:omega}) was typically 32~eV/[10$^{15}$ atoms cm$^{-2}$] and varied by up to 10\% from target to target and depending on degradation. The uncertainty in the stopping power value for each individual target was as quoted in Table \ref{tab:unc}.

Our new value of the $E_{\rm R} = 64.5$~keV resonance strength is about a factor of 2 higher than reported in Ref. \cite{blackmon1995}. The reason for such a discrepancy is not obvious. Correcting for the electron screening effect (enhancement factor $f = 1.15$ based on the adiabatic approximation \cite{assenbaum1987,bracci1990}), we obtain a (bare) value $\omega\gamma = 8.7 \pm 1.2_{\rm stat} \pm 0.6_{\rm syst}$~neV.  
From our new resonance strength values, we derive a proton width for the $E_{\rm R} = 64.5$~keV resonance state ($E_{\rm x} = 5672$~keV in $^{18}$F) of $\Gamma_{\rm p} = 40 \pm 6_{\rm stat} \pm 3_{\rm syst}$~neV (screened) and $\Gamma_{\rm p} = 35 \pm 5_{\rm stat} \pm 3_{\rm syst}$~neV (bare), using the approximation $\Gamma_{\rm tot} \sim \Gamma_\alpha = 130 \pm 5$~eV \cite{mak1980}. Our bare $\Gamma_{\rm p}$ value was used to calculate a revised astrophysical rate for the \one\ reaction, using the on-line RATESMC tool \cite{starlib} and modifying the input file employed in \cite{buckner2015}. The proton partial width was the only parameter that we changed in the file.
\begin{table}
\caption{Sources of uncertainty in the determination of the $E_{\rm R} = 64.5$~keV resonance strength. The tail asymmetry error arises from the non-symmetric shape of the alpha particle peak. See Ref. \cite{bruno2015} for details.}
	\begin{tabular}{l c}
	\hline
	Source & Estimated error \\ \hline
	Statistical & $\pm $14.2\% \\
	Efficiency & $\pm $5.5\% \\
	Stopping power & $\pm$4.0\% \\
	Charge integration & $\pm$2.0\% \\
	Tail asymmetry & +2.0\%\\
	\hline
	\end{tabular}
\label{tab:unc}
\end{table}

\begin{figure}[th]
\resizebox{0.5\textwidth}{!}{
 \includegraphics{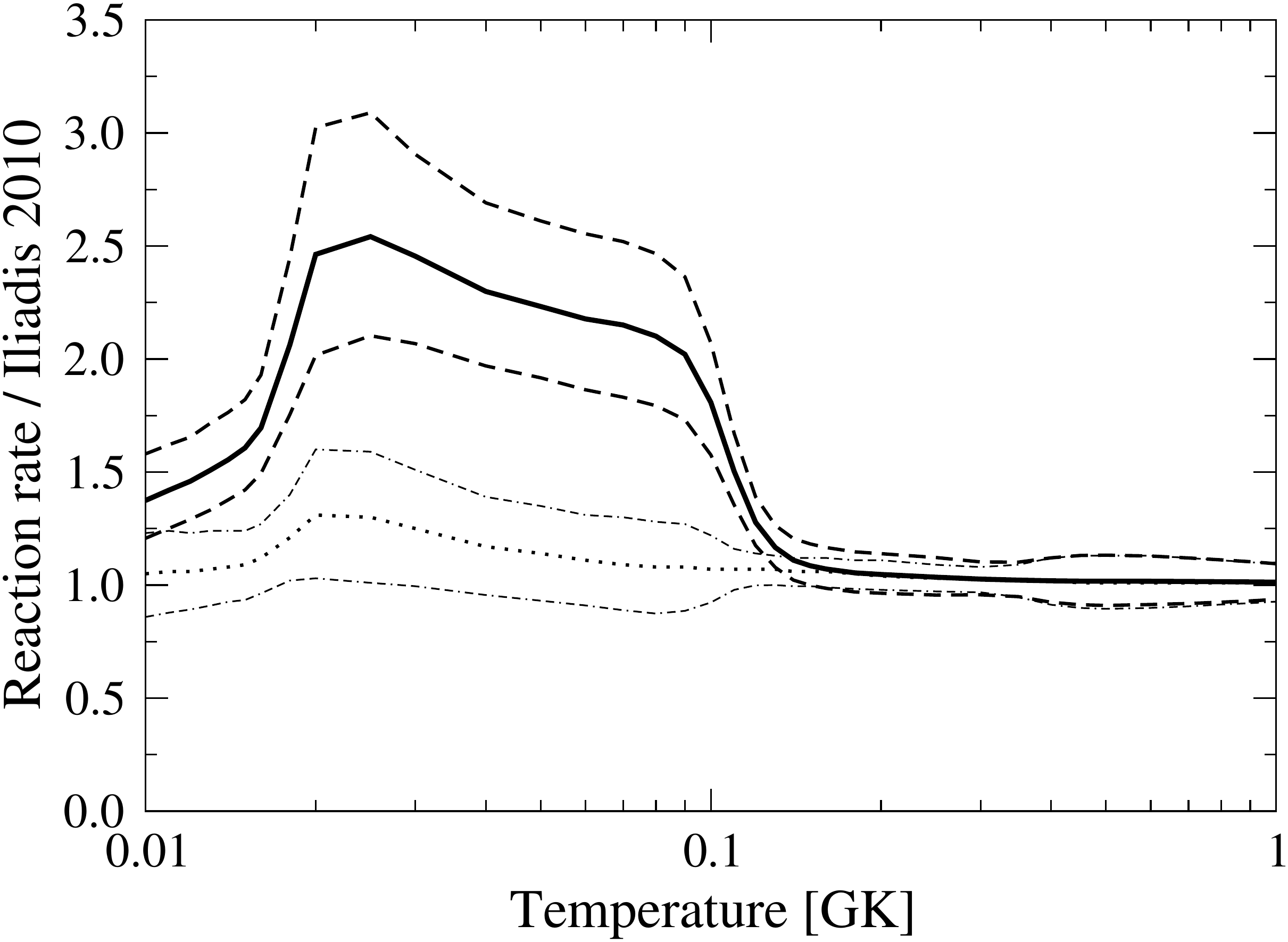}
}
\caption{Ratio of the \one\ reaction rate of the present work (solid line) and of Buckner \cite{buckner2015} (dotted line) to the \one\ reaction rate of Iliadis \cite{iliadis2010}. Dashed and dotted-dashed lines correspond to upper and lower limits as given here and in Buckner {\em et al.}, respectively.}
\label{fig:rate}
\end{figure}

Our new recommended rate is presented in Table \ref{tab:rate} and a comparison with the rate of Ref. \cite{iliadis2010} (widely used in recent astrophysical models) is shown in Fig. \ref{fig:rate}. 
Note that our new recommended reaction rate is a factor of 2 higher at temperatures of astrophysical interest for AGB stars, entirely due to our enhanced value of the $E_{\rm R} = 64.5$~keV resonance strength. 
We do not find evidence of any increase in the rate due to the -2~keV sub-threshold state in $^{18}$F (as claimed in \cite{buckner2015}) and conclude that the 30\% increase reported in \cite{buckner2015} is chiefly due to a revision in the most recent value of the proton separation energy $S_{\rm p} = 5607.1 \pm 0.5$~keV \cite{wang2012} used here and in \cite{buckner2015} as compared to the value $S_{\rm p} = 5606.5 \pm 0.5$~keV \cite{audi-wapstra2003} previously adopted by Iliadis {\em et al.} \cite{iliadis2010}, which leads to $E_{\rm R} = 64.5 \pm 0.5$~keV used in this work. 

The increase in the \one\ reaction rate has important implications for the interpretation of several astronomical observations of the $^{17}$O/$^{16}$O isotopic ratio, including infrared and radio spectra of stars and diffuse matter, as well as abundance measurements in various solar-system components \cite{nittler1997} and in pre-solar stardust inclusions in pristine meteorites \cite{lugaro2007}. The faster destruction rate of $^{17}$O at typical temperatures of shell H-burning translates into a factor of 2 reduction in the equilibrium value of the $^{17}$O/$^{16}$O ratio. However, a full evaluation of the implications of this result on the evolution of the stellar surface composition and corresponding stellar yields requires stellar models with the appropriate treatment of all mixing processes that can bring H-burning ashes to the stellar surface. 
These issues will be addressed in greater detail in forthcoming work \cite{straniero2016,lugaro2016}.
Preliminary calculations for intermediate-mass stars show that the $^{17}$O/$^{16}$O ratio after the first dredge up is up to 20\% smaller than previously estimated, with the maximum variation obtained for stars with $M\sim 2.5$~M$_\odot$. 
Similarly, the $^{17}$O/$^{16}$O ratio is reduced by roughly a factor of 2 in massive asymptotic giant branch (AGB) stars that experience hot bottom burning.

\begin{table}
\caption{Reaction rate $N_A \left<\sigma v\right>$ (cm$^3$ mol$^{-1}$ s$^{-1}$) of the \one\ reaction as a function of temperature (GK). Recommended lower and upper limits are also provided, as calculated according to Ref. \cite{longland2010}.}
	\footnotesize
	\begin{tabular}{ c c c c }
	\hline
T$_9$	&  low	& median & high \\
	\hline
1.00E-02	&	6.89E-25	&	7.87E-25	& 	9.06E-25	\\
1.10E-02	&	7.52E-24	&	8.52E-24	& 	9.77E-24	\\
1.20E-02	&	6.29E-23	&	7.13E-23	& 	8.13E-23	\\
1.30E-02	&	4.26E-22	&	4.81E-22	& 	5.47E-22	\\
1.40E-02	&	2.42E-21	&	2.73E-21	& 	3.10E-21	\\
1.50E-02	&	1.20E-20	&	1.35E-20	& 	1.53E-20	\\
1.60E-02	&	5.43E-20	&	6.16E-20	& 	6.99E-20	\\
1.80E-02	&	1.14E-18	&	1.33E-18	& 	1.58E-18	\\
2.00E-02	&	2.99E-17	&	3.67E-17	& 	4.51E-17	\\
2.50E-02	&	3.03E-14	&	3.66E-14	& 	4.41E-14	\\
3.00E-02	&	3.36E-12	&	3.99E-12	& 	4.76E-12	\\
4.00E-02	&	1.13E-09	&	1.32E-09	& 	1.55E-09	\\
5.00E-02	&	3.41E-08	&	3.99E-08	& 	4.67E-08	\\
6.00E-02	&	3.13E-07	&	3.68E-07	& 	4.30E-07	\\
7.00E-02	&	1.48E-06	&	1.74E-06	& 	2.04E-06	\\
8.00E-02	&	4.64E-06	&	5.48E-06	& 	6.44E-06	\\
9.00E-02	&	1.16E-05	&	1.35E-05	& 	1.58E-05	\\
1.00E-01	&	2.69E-05	&	3.09E-05	& 	3.56E-05	\\
1.10E-01	&	6.90E-05	&	7.65E-05	& 	8.52E-05	\\
1.20E-01	&	1.94E-04	&	2.11E-04	& 	2.30E-04	\\
1.30E-01	&	5.46E-04	&	5.91E-04	& 	6.39E-04	\\
1.40E-01	&	1.42E-03	&	1.54E-03	& 	1.67E-03	\\
1.50E-01	&	3.37E-03	&	3.66E-03	& 	3.97E-03	\\
1.60E-01	&	7.23E-03	&	7.86E-03	& 	8.54E-03	\\
1.80E-01	&	2.59E-02	&	2.82E-02	& 	3.06E-02	\\
2.00E-01	&	7.16E-02	&	7.77E-02	& 	8.44E-02	\\
2.50E-01	&	4.33E-01	&	4.70E-01	& 	5.09E-01	\\
3.00E-01	&	1.60E+00	&	1.72E+00	& 	1.84E+00	\\
3.50E-01	&	5.95E+00	&	6.42E+00	& 	6.90E+00	\\
4.00E-01	&	2.37E+01	&	2.60E+01	& 	2.86E+01	\\
4.50E-01	&	8.35E+01	&	9.28E+01	& 	1.03E+02	\\
5.00E-01	&	2.44E+02	&	2.72E+02	& 	3.04E+02	\\
6.00E-01	&	1.28E+03	&	1.42E+03	& 	1.58E+03	\\
7.00E-01	&	4.21E+03	&	4.64E+03	& 	5.13E+03	\\
8.00E-01	&	1.04E+04	&	1.14E+04	& 	1.25E+04	\\
9.00E-01	&	2.12E+04	&	2.30E+04	& 	2.51E+04	\\
1.00E+00	&	3.78E+04	&	4.08E+04	& 	4.42E+04	\\
	\hline
	\end{tabular}
\label{tab:rate}
\end{table}

In summary, we reported on the most accurate value for the $E_{\rm R} = 64.5$~keV resonance strength in the \one\ reaction from a direct measurement at the underground LUNA accelerator. Major improvements compared to previous direct studies were possible thanks to 
generally improved experimental conditions including: a 15-fold reduction in natural background in the region of interest of the detected alpha particles; a very accurate definition of the region of interest based on the $E_{\rm R} = 183$~keV resonance in \one, not known at the time of previous measurements; and a precise energy calibration [45].
%a 15-fold reduction in background in the energy region of detected alpha particles and to a very accurate definition of the region of interest based on the $E_{\rm R} = 183$~keV resonance in \one. 
Our reaction rate is about a factor of 2 to 2.5 higher than previous rates reported by Iliadis {\em et al.} \cite{iliadis2010} and by Buckner {\em et al.} \cite{buckner2015}. The effect of this revised rate on the nucleosynthesis of $^{17}$O has been briefly presented in relation to hydrogen burning in AGB stars and pre-solar grain compositions. More detailed implications will be reported in forthcoming work.\\

%\subsection{Acknowledgements}

The authors gratefully acknowledge M.Q. Buckner and C. Iliadis for providing the input file used in Ref. \cite{buckner2015}, and M.  Lugaro and V. Paticchio for a critical reading of the manuscript. Financial support by INFN, the Science and Technology Facilities Council, OTKA (Grant No. K108459), and the Helmholtz Association through the Nuclear Astrophysics Virtual Network (NAVI, VH-VI-417) are also gratefully acknowledged. C.G.B. is supported by a Scottish Universities Physics Alliance PhD scholarship. \\

$^*$Corresponding authors: m.aliotta@ed.ac.uk, carlo.bruno@ed.ac.uk

\end{document}